\documentstyle[preprint,aps]{revtex}

\begin{document}
\draft
\preprint{KANAZAWA 94-11, May 1994}
\title{
Monopole action from vacuum configurations in compact QED
}
\author{
Hiroshi Shiba
\footnote{ 
 E-mail address:shiba@hep.s.kanazawa-u.ac.jp}
and Tsuneo Suzuki
\footnote{
 E-mail address:suzuki@hep.s.kanazawa-u.ac.jp
}}
\address{
Department of Physics, Kanazawa University, Kanazawa 920-11, Japan
}
\maketitle
\begin{abstract}
It is possible to derive 
a monopole action from vacuum configurations 
obtained in Monte-Carlo simulations
 extending the method developed by Swendsen. We apply the method to 
compact QED 
both in the Villain and in the Wilson forms. 
The action of the natural monopoles in the Villain case 
is in fairly good agreement with 
that derived by the exact dual transformation. 
Comparing the monopole actions, we find (1) 
the DeGrand-Toussaint monopole definition may be useful 
for $\beta_V $ larger than about 0.5, 
(2) the Villain model well approximates the Wilson one for
 $\beta$ smaller than $\beta_c$ and (3) 
 in the Wilson action  the monopole condensation 
occurs in the confinement phase and $\beta_c$ may be explained 
by the energy-entropy balance of monopole loops
like in the Villain case.
\end{abstract}

\narrowtext
\newpage
To understand confinement mechanism is very important 
but still unresolved 
problem in quantum chromodynamics (QCD). 
A promising idea is that the (dual) 
Meissner effect due to condensation 
of some magnetic
quantity is the color confinement mechanism 
in QCD\cite{thooft1,mandel}.
This picture is realized in the confinement phase of lattice compact
QED\cite{poly,bank,degrand,peskin,frolich,smit,stack}. 

The QED partition function in the Villain form\cite{villain,janke} is
\begin{equation}
Z_V = \prod_{s,\mu} \int_{-\pi}^{\pi}d\theta(s,\mu) \prod_{s,\mu >\nu}
\sum_{n_{\mu\nu}(s)=-\infty}^{\infty} 
\exp(-\frac{\beta_V}{2}\sum_{s,\mu >\nu}
[\theta_{\mu\nu}(s)-2\pi n_{\mu\nu}(s)]^2), \label{villain}
\end{equation}
where the plaquette variable $\theta_{\mu\nu}(s)\in(-4\pi,4\pi)$ is given 
by the link angles $\theta(s,\mu)\in[-\pi,\pi)$ 
as $\theta_{\mu\nu}(s)=\theta(s,\mu)+
\theta(s+\hat\mu,\nu)-\theta(s+\hat\nu,\mu)-\theta(s,\nu)$.
A dual transformation can be done exactly
, leading us to a partition function describing a monopole Coulomb gas
\cite{bank,peskin,frolich,smit}:
\begin{equation}
Z_m = (\prod_{s,\mu}\sum_{m_{\mu}(s)=-\infty}^{\infty})
(\prod_s \delta_{\partial'_{\mu}m_{\mu}(s),0})
\exp( -2\pi^2\beta_V \sum_{s, s',\mu}m_{\mu}(s)D(s-s')m_{\mu}(s')),
 \label{zmono}
\end{equation}
where $D(s-s')$ is the lattice Coulomb propagator and 
$\partial'$ is the backward difference. Here $m_{\mu}(s)$ 
is a monopole current:
\begin{equation}
m_{\mu}(s) = \frac{1}{2}\epsilon_{\mu\nu\rho\sigma}
\partial_{\nu}n_{\rho\sigma}(s+\hat{\mu}), \label{mcur} 
\end{equation}
which is defined on a link of the dual lattice 
(the lattice with origin shifted by 
half a lattice distance all in four directions).
$\partial$ is the forward difference.
Note that the coupling constant in (\ref{zmono}) satisfies the 
Dirac quantization condition between electric and magnetic charges.
One can prove the monopole condensation from energy-entropy balance 
using the action (\ref{zmono}) and the (bulk) 
entropy $\sim$ $L\times $ln7 where $L$ is the monopole loop length
\cite{bank,peskin,frolich,smit}. 
The critical $\beta_V$ is estimated to be 
$\beta_V^c =$ ln7/($2\pi^2 D(0))=.637$ 
\cite{bank} which is in good 
agreement with the Monte-Carlo data $\beta_V^c =.645$\cite{jersak}.

On the other hand, the Wilson partition function of lattice compact QED is
\begin{equation}
Z_W = \prod_{s,\mu} \int_{-\pi}^{\pi}d\theta(s,\mu) 
\exp[ -\beta_W \sum_{s,\mu >\nu} 
(1-cos\theta_{\mu\nu}(s))]. \label{wilson}
\end{equation}
Since the action of the Wilson form is the simplest and 
is used commonly also in non-abelian 
gauge theories, it is important to study the confinement 
mechanism of compact QED 
with this action. 
However the exact dual transformation 
which enables us to make the above analytic study 
is impossible in this case, although 
this action (\ref{wilson}) is easy to perform Monte-Carlo simulations. 
To speak rigorously, hence, the monopole condensation is not yet proved in the 
case of the Wilson action.

Using the Fourier transformation of (\ref{wilson}) and the 
property of the modified Bessel function, the Wilson 
action is approximated for 
very large $\beta$ and for small $\beta$ 
by the Villain form (\ref{villain}).
How good the Villain approximation to the Wilson form is was studied in 
\cite{villain,janke} extensively. 
In Ref.\cite{janke}, physical quantities like internal energy 
and specific heats can be well reproduced for small $\beta_V$ 
up to $\beta_V^c$, if both the coupling constants satisfy 
\begin{equation}
\beta_V^{-1} = -2\rm{ln}(\frac{I_1(\beta_W)}{I_0(\beta_W)}), \label{betawv}
\end{equation}
where $I_0$ and $I_1$ are usual modified Bessel functions.
The approximation becomes rapidly worse above the transition point except 
for very large $\beta$. For moderately large $\beta_V$, the Villain 
partition function gives results rather similar 
to a mixed cosine action with a 
modified coupling relation.
However even if some physical quantities are well reproduced by the Villain 
approximation, it does not always mean that the confinement 
mechanisms of both models are the same. Really it is pointed in \cite{janke} 
that there are physical quantities not well reproduced even for small $\beta$.
Considering applications to QCD, we want to know directly the confinement 
mechanism in the Wilson action. 

For that purpose, one has to find first monopole currents in the Wilson 
action where there is no natural definition like (\ref{mcur})
contrary to the Villain case. The only known definition is that 
given by DeGrand and Toussaint\cite{degrand}. The plaquette variable 
$\theta_{\mu\nu}(s)$ can be decomposed into 
\begin{equation}
\theta_{\mu\nu}(s)= \bar{\theta}_{\mu\nu}(s) +2\pi m_{\mu\nu}(s),
\end{equation}
where $\bar{\theta}_{\mu\nu}(s)\in [-\pi,\pi)$ and $m_{\mu\nu}(s)$ 
can be regarded as a number of the Dirac string penetrating the plaquette.
DeGrand and Toussaint\cite{degrand} defined a monopole current as 
\begin{equation}
m_{\mu}^{DG}(s) = \frac{1}{2}\epsilon_{\mu\nu\rho\sigma}
\partial_{\nu}m_{\rho\sigma}(s+\hat{\mu}). \label{dgcur} 
\end{equation}
Recently Schram and Teper\cite{schram}
 studied how well the DGT monopoles 
$m_{\mu}^{DG}(s)$ approximate the natural ones $m_{\mu}(s)$ based on Monte-Carlo 
simulations of the Villain form. 
They suggested the DGT prescription can be 
meaningfully used in the studies of the U(1) phase transition.

Next we have to find a partition function described in terms of the monopole 
currents corresponding to the dual form of the Wilson action 
in order to study the energy entropy balance. 
Actually an exact dual transformation is impossible 
in almost all models where dual variables are expected to play an important 
role in the dynamics. It is very important to develop a method determining 
a dual theory for such models.
It is the most important original point of this note to 
give a method fixing a monopole action from a given 
ensemble of configurations generated in  Monte-Carlo simulations.
To demonstrate the usefulness of our method, 
we first apply the method to the natural monopoles in the Villain 
case of $U(1)$ lattice gauge 
theory where the exact monopole partition function is 
known as in (\ref{zmono}). 
Next we derive the action for the DGT monopoles in the Villain case 
and test reliability of the DGT prescription. Finally we apply the method
to the Wilson case to check the occurence of the monopole condensation.
A brief description of our method and preliminary applications were
 given also in \cite{shiba1,shiba3}.

We extend the method of Swendsen\cite{swendsen} to 
determine an action from 
a given ensemble of monopole loops in vacuum configurations. 
A theory of monopole loops is given in general by 
the following partition 
function  
\begin{equation}
Z= (\prod_{s,\mu}\sum_{k_{\mu}(s)=-\infty}^{\infty})
(\prod_s \delta_{\partial'_{\mu}k_{\mu}(s),0})
\exp(-S[k]), \label{zmono2}
\end{equation}
where $k_{\mu}(s)$ 
is the conserved integer-valued monopole current defined above in
(\ref{mcur}) or (\ref{dgcur}) and $S[k]$ is a monopole action describing 
the theory. 
Consider a set of all independent operators which are summed up 
over the whole lattice. We denote each operator as $S_i [k]$. Then 
the action can be written as a linear combination of these operators:
\begin{equation}
S[k] = \sum_i f_i S_i [k], 
\end{equation}
where $f_i$ are coupling constants. 
Really the monopole 
partition function (\ref{zmono}) takes this form. 
The expectation value of an operator of monopole currents are estimated as
\begin{eqnarray}
\langle O[k]\rangle =
\frac{
(\prod_{s,\mu}\sum_{k_{\mu}(s)=-\infty}^{\infty})
(\prod_s \delta_{\partial'_{\mu}k_{\mu}(s),0}) O[k]
\exp(-\sum_{i}f_i S_i [k])}
{(\prod_{s,\mu}\sum_{k_{\mu}(s)=-\infty}^{\infty})
(\prod_s \delta_{\partial'_{\mu}k_{\mu}(s),0})
\exp(-\sum_{i}f_i S_i [k])}. \label{ok}
\end{eqnarray}

Let us now determine the monopole action using the monopole 
current ensemble which are calculated  by vacuum 
configurations generated by Monte-Carlo simulations.
Since the dynamical variables here are $k_{\mu}(s)$ satisfying the 
conservation rule, it is necessary to extend the 
original Swendsen method. 
Consider a plaquette $(s',\hat{\mu'},\hat{\nu'})$ instead of a link 
on the dual lattice.
Define $\hat{S}_i [k]$ as a part of an operator $ S_i [k]$ containing 
the monopole currents on the links in the plaquette chosen, i.e., 
$k_{\mu'}(s'), k_{\nu'}(s'+\hat{\mu'}), k_{\mu'}(s'+\hat{\nu'})$ and 
$k_{\nu'}(s')$. Then we get 
\begin{eqnarray}
& &(\prod_{s,\mu}\sum_{k_{\mu}(s)=-\infty}^{\infty})
(\prod_s \delta_{\partial'_{\mu}k_{\mu}(s),0}) S_a [k]
\exp(-\sum_i f_i S_i [k]) \nonumber \\
& = &(\prod_{s,\mu}^{\ \ \ \prime}
\sum_{k_{\mu}(s)=-\infty}^{\infty})
(\prod_s^{\ \ \ \prime}\delta_{\partial'_{\mu}k_{\mu}(s),0}) 
 [\sum_{k_{\mu'}(s')=-\infty}^{\infty}
\sum_{k_{\nu'}(s'+\hat{\mu'})=-\infty}^{\infty}
\sum_{k_{\mu'}(s'+\hat{\nu'})=-\infty}^{\infty}
\sum_{k_{\nu'}(s')=-\infty}^{\infty} \nonumber \\
& & \delta_{\partial'_{\mu}k_{\mu}(s'),0}
\delta_{\partial'_{\mu}k_{\mu}(s'+\hat{\mu'}),0}
\delta_{\partial'_{\mu}k_{\mu}(s'+\hat{\nu'}),0}
\delta_{\partial'_{\mu}k_{\mu}(s'+\hat{\mu'}+\hat{\nu'}),0}
\ S_a [k] \exp(-\sum_i f_i \hat{S}_i [k])] \nonumber \\
&  & \exp(-\sum_i f_i (S_i [k]- \hat{S}_i [k])) 
, \label{z1}
\end{eqnarray}
where the primed product means removing the sites and the links 
in the plaquette chosen.
Note here that the sum of the current conservations on the four sites 
$\partial'_{\mu}k_{\mu}(s')+
\partial'_{\mu}k_{\mu}(s'+\hat{\mu'})+
\partial'_{\mu}k_{\mu}(s'+\hat{\nu'})+
\partial'_{\mu}k_{\mu}(s'+\hat{\mu'}+\hat{\nu'})$
does not contain any current on the four links of the plaquette adopted.
Hence 
\begin{eqnarray}
(\ref{z1}) & = &
(\prod_{s,\mu}^{\ \ \ \prime}\sum_{k_{\mu}(s)=-\infty}^{\infty})
(\prod_s^{\ \ \ \prime}\delta_{\partial'_{\mu}k_{\mu}(s),0})
\delta_{
\partial'_{\mu}k_{\mu}(s')+
\partial'_{\mu}k_{\mu}(s'+\hat{\mu'})+
\partial'_{\mu}k_{\mu}(s'+\hat{\nu'})+
\partial'_{\mu}k_{\mu}(s'+\hat{\mu'}+\hat{\nu'}),0} \nonumber \\
&   &  \exp(-\sum_i f_i (S_i [k]- \hat{S}_i [k])) 
(\sum\delta)_{k}S_a [k]\exp(-\sum_i f_i \hat{S}_i [k])\\
& = & 
(\prod_{s,\mu}\sum_{k_{\mu}(s)=-\infty}^{\infty})
(\prod_s \delta_{\partial'_{\mu}k_{\mu}(s),0})
\bar{S}_a [k]\exp(-\sum_i f_i S_i [k])
,\label{z2}
\end{eqnarray}
where 
\begin{eqnarray}
\bar{S}_a [k]\equiv
\frac{(\sum\delta)_{\hat{k}} S_a [\hat{k}, \{k\}']
\exp(-\sum_{i}f_i\hat{S}_i [\hat{k},\{k\}'])}
{(\sum\delta)_{\hat{k}} \exp(-\sum_{i}f_i\hat{S}_i [\hat{k},\{k\}'])}.
\end{eqnarray}
Here 
$\{k\}'$ do not contain the four currents on the 
links of the plaquette considered and
\begin{eqnarray}
(\sum\delta)_{\hat{k}}
& \equiv &
\sum_{\hat{k}_{\mu'}(s')=-\infty}^{\infty}
\sum_{\hat{k}_{\nu'}(s'+\hat{\mu'})=-\infty}^{\infty}
\sum_{\hat{k}_{\mu'}(s'+\hat{\nu'})=-\infty}^{\infty}
\sum_{\hat{k}_{\nu'}(s')=-\infty}^{\infty}\nonumber \\
&  &\delta_{\partial'_{\mu}\hat{k}_{\mu}(s'),0}
\delta_{\partial'_{\mu}\hat{k}_{\mu}(s'+\hat{\mu'}),0}
\delta_{\partial'_{\mu}\hat{k}_{\mu}(s'+\hat{\nu'}),0}. 
\end{eqnarray}
Since there are current conservations at all sites in (\ref{z2}), 
we get 
\begin{eqnarray}
\partial'_{\mu}\hat{k}_{\mu}(s') & = & \partial'_{\mu}k_{\mu}(s') 
+\hat{k}_{\mu'}(s')+\hat{k}_{\nu'}(s')-k_{\mu'}(s')-k_{\nu'}(s')\\
& = & \hat{k}_{\mu'}(s')+\hat{k}_{\nu'}(s')-k_{\mu'}(s')-k_{\nu'}(s').
\end{eqnarray}
Similarly, we see 
\begin{eqnarray}
\partial'_{\mu}\hat{k}_{\mu}(s'+\hat{\mu'}) = 
 \hat{k}_{\nu'}(s'+\hat{\mu'})-\hat{k}_{\mu'}(s')
-k_{\nu'}(s'+\hat{\mu'})+k_{\mu'}(s'),\\
\partial'_{\mu}\hat{k}_{\mu}(s')+\partial'_{\mu}\hat{k}_{\mu}(s'+\hat{\nu'})= 
 \hat{k}_{\mu'}(s'+\hat{\nu'})+\hat{k}_{\mu'}(s')
-k_{\mu'}(s'+\hat{\nu'})-k_{\mu'}(s').
\end{eqnarray}
Now use the following relation:
\begin{equation}
\sum_{M=-\infty}^{\infty} \delta_{\hat{k}_{\mu'}(s'),k_{\mu'}(s')+M} =1.
\label{dkm}
\end{equation}
Then 
\begin{eqnarray}
& & (\sum\delta)_{\hat{k}}
S_a [\hat{k}_{\mu'}(s'), \hat{k}_{\nu'}(s'+\hat{\mu'}),
\hat{k}_{\mu'}(s'+\hat{\nu'}), \hat{k}_{\nu'}(s'), \{k\}'] 
 = \sum_{M=-\infty}^{\infty} 
\sum_{\hat{k}_{\mu'}(s')=-\infty}^{\infty}
\delta_{\hat{k}_{\mu'}(s'),k_{\mu'}(s')+M}\nonumber \\
& \times & 
\sum_{\hat{k}_{\nu'}(s'+\hat{\mu'})=-\infty}^{\infty} 
\delta_{\hat{k}_{\nu'}(s'+\hat{\mu'}),
k_{\nu'}(s'+\hat{\mu'})+M}
\sum_{\hat{k}_{\mu'}(s'+\hat{\nu'})=-\infty}^{\infty} 
\delta_{\hat{k}_{\mu'}(s'+\hat{\nu'}),
k_{\mu'}(s'+\hat{\nu'})-M}
\sum_{\hat{k}_{\nu'}(s')=-\infty}^{\infty}
\delta_{\hat{k}_{\nu'}(s'),k_{\nu'}(s')-M}\nonumber \\
& \times  &
S_a [k_{\mu'}(s')+M, k_{\nu'}(s'+\hat{\mu'})+M,
k_{\mu'}(s'+\hat{\nu'})-M, k_{\nu'}(s')-M, \{k\}']\\
 & = & \sum_{M=-\infty}^{\infty} 
S_a [k_{\mu'}(s')+M, k_{\nu'}(s'+\hat{\mu'})+M,
k_{\mu'}(s'+\hat{\nu'})-M, k_{\nu'}(s')-M, \{k\}'].  
\end{eqnarray}
When the DGT monopoles are used, the sum with respect to $M$ is restricted 
from the minimum $m_1$ to the maximum $m_2$, where
\begin{eqnarray}
m_1 = -2-\rm{Min}\{k_{\mu'}(s'), k_{\nu'}(s'+\hat{\mu'}),
-k_{\mu'}(s'+\hat{\nu'}), -k_{\nu'}(s')\},\\
m_2 = 2-\rm{Max}\{k_{\mu'}(s'), k_{\nu'}(s'+\hat{\mu'}),
-k_{\mu'}(s'+\hat{\nu'}), -k_{\nu'}(s')\}.
\end{eqnarray}
Note that the DGT monopoles take integer values between -2 and 2 only 
as seen from the definition.
Then using (\ref{ok}), we get 
\begin{eqnarray}
\langle S_a [k]\rangle = \langle \bar{S}_a [k]\rangle, \label{sbar} 
\end{eqnarray}
where 
\begin{eqnarray}
\bar{S}_a [k] = \frac{\sum_{M=-\infty}^{\infty} S_a[\bar{k}]
\exp(-\sum_{i}f_i \hat{S}_i [\bar{k}])}
{\sum_{M=-\infty}^{\infty} \exp(-\sum_{i}f_i \hat{S}_i [\bar{k}])}
\end{eqnarray}
and 
\begin{equation}
\bar{k}_{\mu}(s)\equiv k_{\mu}(s) + M(
\delta_{s,s'}\delta_{\mu,\mu'}
+\delta_{s,s'+\hat{\mu'}}\delta_{\mu,\nu'}
-\delta_{s,s'+\hat{\nu'}}\delta_{\mu,\mu'}
-\delta_{s,s'}\delta_{\mu,\nu'}).\label{kbar}
\end{equation}

Introducing a new set of coupling constants 
$ \{\tilde{f}_i\} $, we define
\begin{equation}
\tilde{S}_a [k]= \frac{\sum_{M=-\infty}^{\infty}S_a[\bar{k}]
\exp(-\sum_i \tilde{f}_i \hat{S}_i [\bar{k}])}
{\sum_{M=-\infty}^{\infty}\exp(-\sum_i \tilde{f}_i \hat{S}_i [\bar{k}])},
\label{stilde}
\end{equation}
where $\bar{k}$ is defined in (\ref{kbar}).
When all $\tilde{f}_i$ are equal to $ f_i $,
one can prove an equality   from (\ref{sbar})
$$ \langle \tilde{S}_a [k] \rangle  =  \langle S_a [k] \rangle. $$
When there are some $\tilde{f}_i$ not equal to $ f_i $,
 one may expand the difference  as follows:  
\begin{equation}
\langle \tilde{S}_a - S_a \rangle = 
\sum_b 
\langle \widetilde{S_a S_b}-\tilde{S}_a\tilde{S}_b \rangle
(f_b - \tilde{f}_b)    \label{sdif}, 
\end{equation}
where only the first order terms up to $O(f_b -\tilde{f}_b)$ 
are written down.
This allows an iteration scheme for determination 
of the unknown constants
$f_i$ from the ensemble of $\{k_{\mu}(s)\}$, which are generated in 
Monte-Carlo simulations. 

Practically we have to restrict the number of 
interaction terms
\footnote{
All possible types of interactions are not independent, since 
$\partial'_{\mu}k_{\mu}(s)=0$. We can get rid of almost 
all interactions between 
different components of the currents from the quadratic action 
by use of the conservation rule.
}. 
We  adopted
quadratic interactions of up to 32 types listed in Table \ref{table1} 
in these 
studies. The first six terms are shown also 
graphically in Fig.\ \ref{action}.

Before getting the configurations, we checked 
the auto-correlation time and thermalization. For thermalization, 
we need more than 500 iterations in the Villain case, whereas 50 iterations
are enough in the Wilson case near the critical $\beta$. The auto-correlation 
quickly disappears in the Wilson case, whereas the situations 
in the Villain case are much worse near the critical $\beta$. Hence
using the Villain action, we generated 100 gauge field configurations 
separated by 50 sweeps after a thermalization of 1000 sweeps 
for $\beta_V < 0.56$.  At $\beta_V =0.62$, we separated 300 iterations.
For more than the critical $\beta_V^c \sim 0.64$, we performed 100 independent 
runs with different initial values and adopted one configuration after 
1000 thermalization loops for each run. Each run in the Villain case was done  
on $8^4$ lattice. In the Wilson case, we took 100 configurations 
separated by  50 sweeps
after 1000 thermalization sweeps for every $\beta_W$ on $8^4$ and $12^4$ 
lattices. The monopole currents are defined following (\ref{mcur}) and
(\ref{dgcur}).
The statistical errors were estimated 
with the jackknife method. 

Our results are the following.
\begin{enumerate}
\item 
The first coupling constants $f_1 \sim f_6$ of the action for the 
natural monopoles  in the Villain case 
are plotted in Fig.\ \ref{villain1} in 
comprison with the theoretical values given by (\ref{zmono}). 
$f_1$ agree well with those of the theoretical values, whereas 
there are small discrepancies with respect to $f_2 \sim f_6$. 
The discrepancies may come from the truncation of the terms of the action 
taken. At the critical $\beta_V^c \sim 0.64$, $f_1$ crosses 
the ln7 entropy line. 

\item 
The same coupling constants of the action for the DGT 
monopoles  in the Villain case 
are plotted in Fig.\ \ref{villain2}. Both $f_1$ of the natural 
and the DGT monopole actions are compared in Fig.\ \ref{villain3}.
We see there is a large deviation for small $\beta$. For $\beta_V > 0.5$, 
we may use the DGT prescription. Schram and Teper\cite{schram} recently showed 
the difference of the natural and the DGT monopoles can be reproduced by 
a random distribution of dipoles having trivial long-distance behaviors.
They concluded therefore the DGT prescription can be used for 
$\beta_V > 0.3$. $f_1$ is the dominant part of the action and it plays an 
important role in the energy-entropy balance. There is a large gap with 
respect to $f_1$ at $\beta_V= 0.3$. It seems dangerous to use the DGT 
definition for such small $\beta_V$. Fortunately the DGT prescription 
looks rather good around the ctitical $\beta_V^c$. 

\item
Using the DGT definition, we get the monopole action in the Wilson case.
The first six coupling constants of the action on $12^4$ lattice 
are plotted in Fig.\ \ref{wilson1}. There is a small volume dependence 
for $\beta_W$ larger than the critical value( $\sim 1.0$). $f_1$ on $8^4$ 
and $12^4$ lattices are compared in Fig.\ \ref{wilson2}. 
The energy of a monopole loop of length $L$ may be approximated by a 
self-energy part $f_1 L$ when $L$ is large\cite{bank,bode}.
The entropy line ln7 and the $f_1$ line 
crosses at $\beta_W = 1.05$ which is very near to the critical 
$\beta_W \sim 1.0$. 
Namely the critical coupling constant determined from the 
monopole condensation due to the energy-entropy balance  of monopole loops 
agrees to that of the deconfinement transition. The monopole 
condensation really occurs in the confinement phase also in the Wilson 
model. This is the first direct evidence for the occurence of the monopole 
condensation in the Wilson action without 
the use of the Villain approximation.
\item 
Similarity of the Villain and the Wilson partition functions is 
studied by comparing both $f_1$ in Fig.\ \ref{wilson3},
 where $\beta_W$ is transformed into $\beta_V$ using (\ref{betawv}).
Both are in good agreement for small $\beta$ less than the critical 
value $\sim 0.64$. For larger $\beta$, both begin to deviate rapidly. This is 
consistent with the conclusion given in \cite{janke}.

\end{enumerate}

Finally we have seen that our method is very useful in the study 
of compact QED. The method can be applied also 
to more interesting QCD cases. For preliminary reports, see 
\cite{shiba1,shiba3}. 
The detail of the results will be published 
elsewhere\cite{shiba4}.

We wish to acknowledge Yoshimi Matsubara for 
useful discussions.
This work is financially supported by JSPS Grant-in Aid for 
Scientific  Research (B)(No.06452028).

\begin{table}
\caption{
The quadratic terms of the monopole action adopted. 
Only the partner of the current multiplied by $k_{\mu}(s)$ 
are listed. All terms in which 
the relation of the two currents is equivalent are added 
to make each $S_i [k]$ invariant under translation and rotation. 
Here 
$\hat{a} \equiv \hat{\nu}+\hat{\rho}$, 
$\hat{b} \equiv \hat{\nu}+\hat{\rho}+\hat{\omega}$,
$\hat{c} \equiv \hat{\mu}+\hat{\nu}+\hat{\rho}+\hat{\omega}$,
$\hat{d} \equiv 2\hat{\mu}+\hat{\nu}$,
$\hat{e} \equiv 2\hat{\nu}+\hat{\mu}$ and
$\hat{f} \equiv 2\hat{\nu}+\hat{\omega}$,
where $\hat{\mu}, \hat{\nu}, \hat{\rho}$ and $\hat{\omega}$
denote unit vectors in four different directions.
\label{table1}
}
\begin{tabular}{rcrcrc}
 i   &   current partner   & i   &   current partner   & i   &   current partner   \\
\tableline
 1&$k_{\mu}(s)                    $&13&$k_{\mu}(s+\hat{e})           $&25&$k_{\mu}(s+3\hat{\mu}+\hat{a})$\\
 2&$k_{\mu}(s+\hat{\mu})          $&14&$k_{\mu}(s+2\hat{\nu})        $&26&$k_{\nu}(s+\hat{d}+2\hat{\omega})$\\
 3&$k_{\mu}(s+\hat{\nu})          $&15&$k_{\mu}(s+\hat{d}+\hat{a})   $&27&$k_{\mu}(s+\hat{d}+\hat{b})  $\\
 4&$k_{\mu}(s+\hat{\mu}+\hat{\nu})$&16&$k_{\mu}(s+3\hat{\mu})        $&28&$k_{\mu}(s+3\hat{\mu}+\hat{b})  $\\
 5&$k_{\mu}(s+\hat{a})            $&17&$k_{\mu}(s+\hat{e}+\hat{\rho})$&29&$k_{\mu}(s+\hat{e}+2\hat{\rho})$\\
 6&$k_{\mu}(s+2\hat{\mu})         $&18&$k_{\mu}(s+\hat{f})           $&30&$k_{\mu}(s+\hat{f}+\hat{\omega})$\\
 7&$k_{\mu}(s+\hat{\mu}+\hat{a})  $&19&$k_{\mu}(s+\hat{d}+\hat{\nu}) $&31&$k_{\mu}(s+\hat{e}+2\hat{\mu})  $\\
 8&$k_{\mu}(s+\hat{b})            $&20&$k_{\mu}(s+\hat{d}+\hat{\mu}) $&32&$k_{\nu}(s+\hat{e}+2\hat{\mu}) $\\
 9&$k_{\mu}(s+\hat{c})            $&21&$k_{\nu}(s+\hat{d}+\hat{\mu}) $&  &$  $\\
10&$k_{\mu}(s+\hat{d})            $&22&$k_{\mu}(s+\hat{c}+\hat{\nu}) $&  &$  $\\
11&$k_{\nu}(s+\hat{d})            $&23&$k_{\mu}(s+\hat{f}+\hat{\rho})$&  &$  $\\
12&$k_{\mu}(s+\hat{d}+\hat{\rho}) $&24&$k_{\mu}(s+2\hat{\mu}+\hat{b})$&  &$  $\\
\end{tabular}
\end{table}
   
\begin{figure}[tb]
\caption{
The first six terms of the monopole action adopted.}
\label{action}
\end{figure}

\begin{figure}[tb]
\caption{
Coupling constants $f_i$ versus $\beta$ in the Villain model
 when the natural monopole currents $m_{\mu}(s)$ are used. The solid (dotted,
 dashed and dott-dashed ) line denotes the theoretical curve of $f_1 
(f_2 =f_3 , f_4 =f_5$ and $f_6$).} 
\label{villain1}
\end{figure}

\begin{figure}[tb]
\caption{
Coupling constants $f_i$ versus $\beta$ in the Villain model
 when the Degrand-Toussaint monopole currents $m_{\mu}^{DG}(s)$ are used.
The lines show the theoretical curves predicted by the monopole 
action for the natural monopoles.} 
\label{villain2}
\end{figure}

\begin{figure}[tb]
\caption{
Coupling constants $f_1$ versus $\beta$ in the Villain model
 for both the natural and the Degrand-Toussaint monopole currents.}
\label{villain3}
\end{figure}

\begin{figure}[tb]
\caption{
Coupling constants $f_i$ versus $\beta$ in the Wilson model on $12^4$ lattice
 where the Degrand-Toussaint monopole currents $m_{\mu}^{DG}(s)$ are used.} 
\label{wilson1}
\end{figure}

\begin{figure}[tb]
\caption{
Coupling constants $f_1$ versus $\beta$ in the Wilson model both 
on $8^4$ and $12^4$ lattices.}
\label{wilson2}
\end{figure}

\begin{figure}[tb]
\caption{
Coupling constants $f_1$ versus $\beta$ in the Villain 
and the Wilson models, where the DGT monopoles are used. The solid line 
denotes the theoretical value of $f_1$ of the action for the natural 
monopoles.
}
\label{wilson3}
\end{figure}

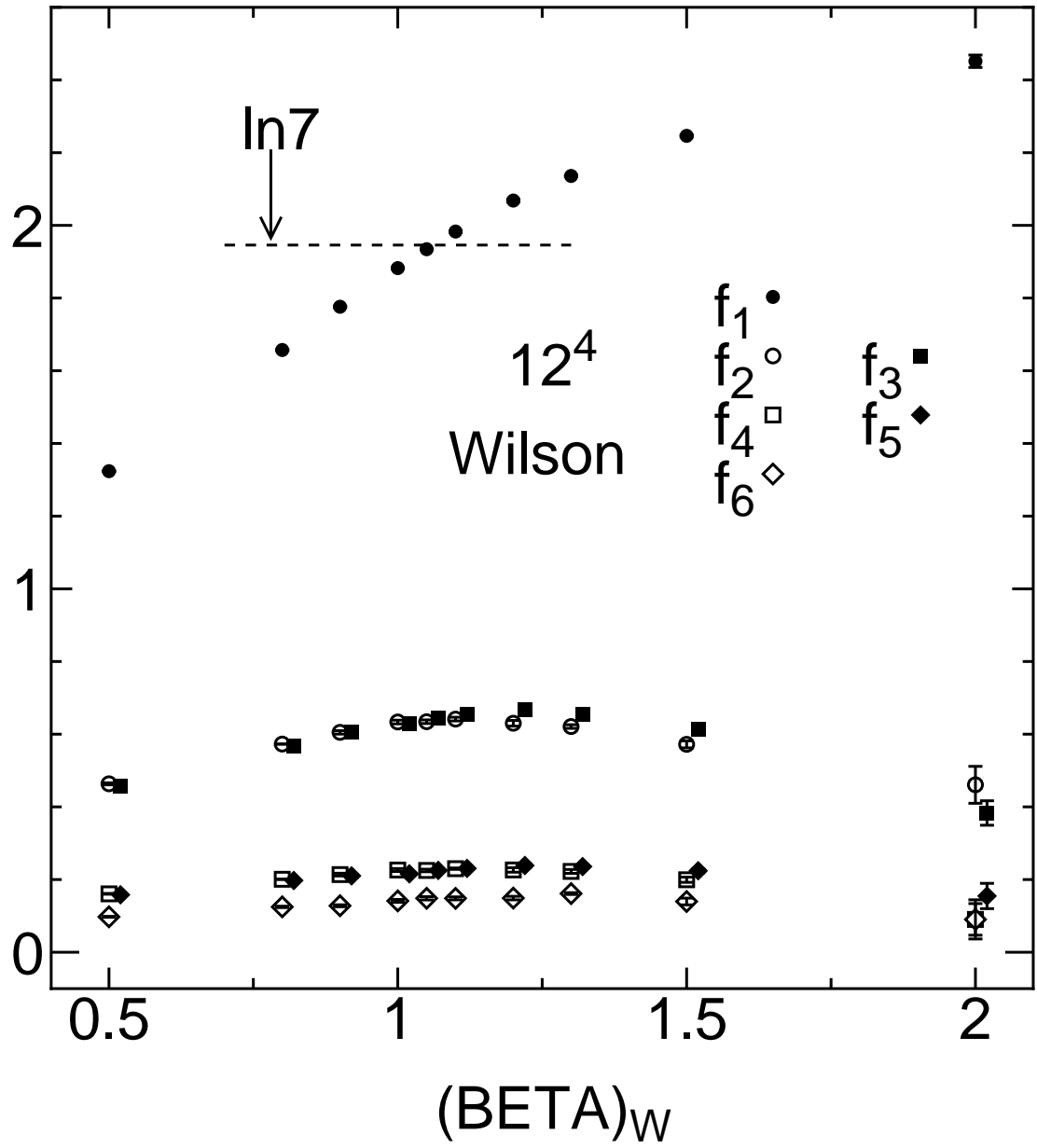
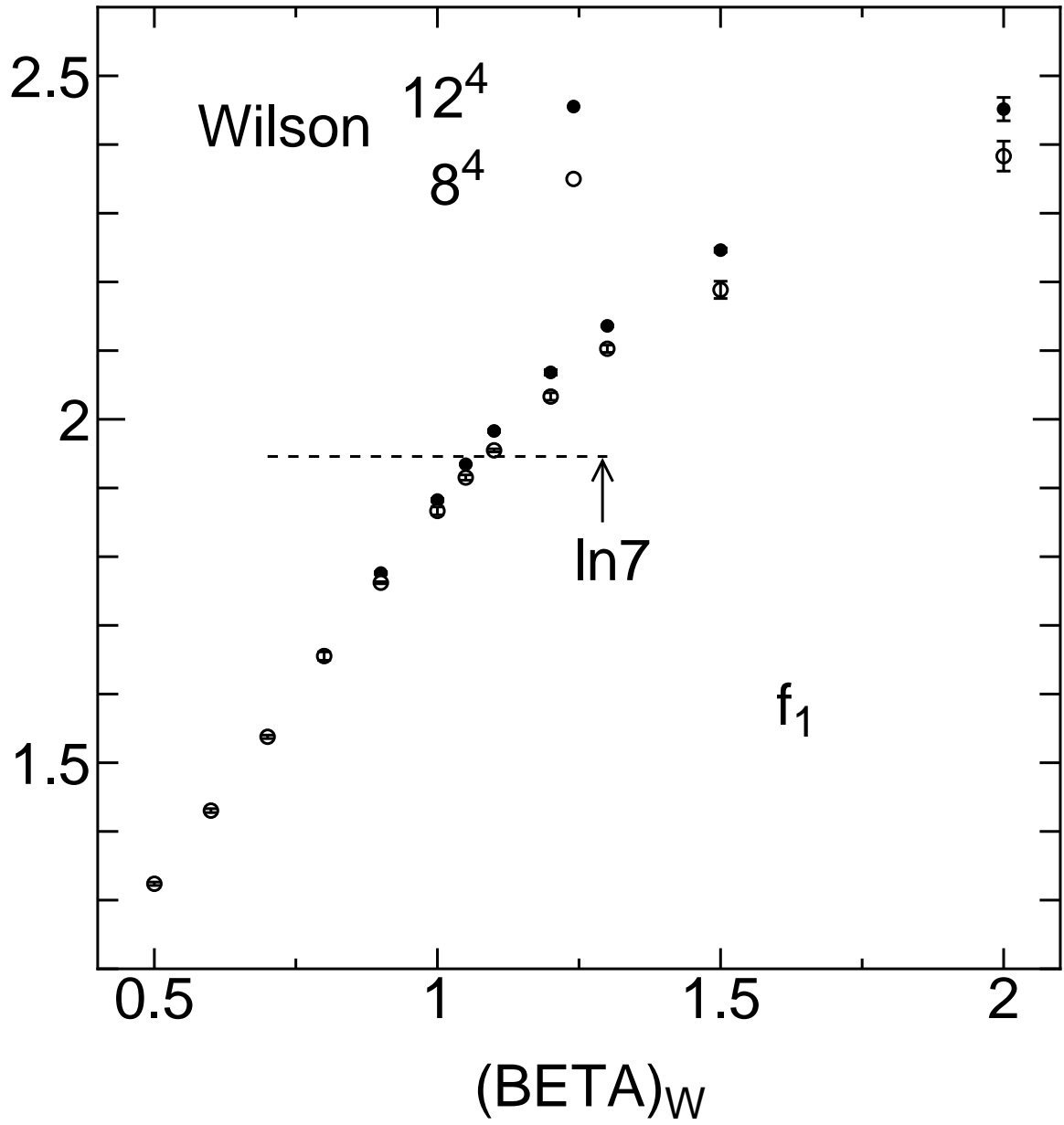
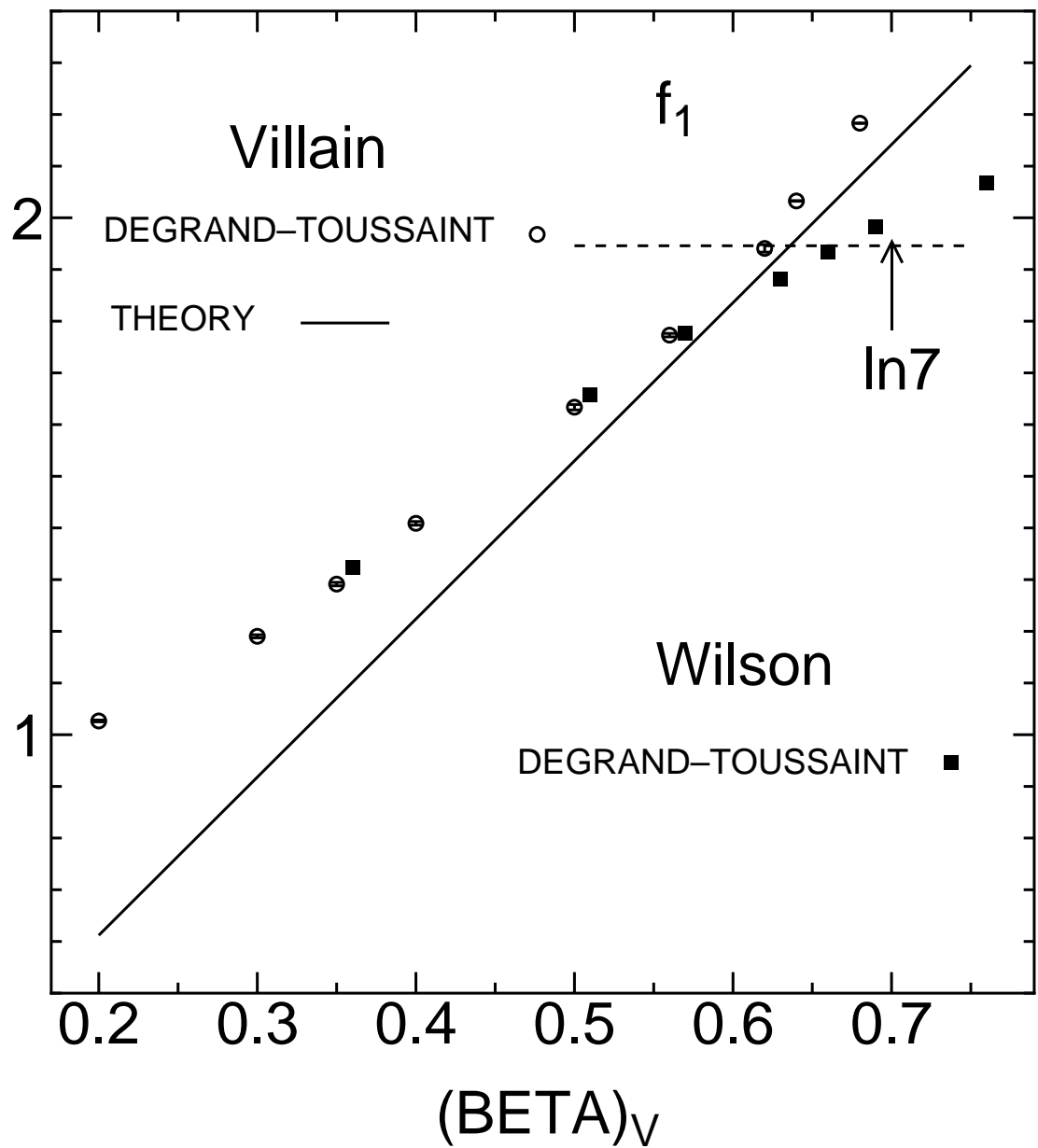
\end{document}